\documentclass[12pt, preprint]{aastex}

\slugcomment{ApJ Letters Accepted}

\newcommand{\Msun}{M$_{\odot}$}

\newcommand{\SI}{[SII]$\lambda$6731}
\newcommand{\Ox}{[OI]$\lambda$6300}

\newcommand{\Ha}{H$\alpha$}

\newcommand{\NII}{[NII]$\lambda$6583}
\newcommand{\OI}{[OI]$\lambda\lambda$6300,6363}

\newcommand{\km}{kms$^{-1}$}

\begin{document}


\title{Discovery of a Bipolar Outflow from 2MASSW J1207334-393254 a 24 M$_{\rm jup}$ Brown Dwarf}


\author{E.T. Whelan\altaffilmark{1}}

\author{T.P. Ray\altaffilmark{1}}

\author{S. Randich\altaffilmark{2}}

\author{F. Bacciotti\altaffilmark{3}}

\author{R. Jayawardhana\altaffilmark{2}}

\author{L. Testi\altaffilmark{2}}

\author{A. Natta\altaffilmark{2}}

\author{S. Mohanty\altaffilmark{4}}


\altaffiltext{1}{School of Cosmic Physics, Dublin Institute for Advanced Studies}
\altaffiltext{2}{Osservatorio Astrofisico di Arcetri}
\altaffiltext{3}{Department of Astronomy and Astrophysics, University of Toronto}
\altaffiltext{4}{Harvard-Smithsonian CfA}


\begin{abstract}

The 24 M$_{\rm jup}$ brown dwarf 2MASS1207-3932 has for some time
 been known to show clear signs of classical T Tauri-like
  accretion. Through analysis of its oxygen forbidden emission we have 
  discovered that it is driving a bipolar outflow. Blue and 
  red-shifted components to the \Ox\ forbidden emission line are seen
   at velocities of - 8\km\ and +4\km\  (on either side of the systemic velocity)
. Spectro-astrometry recovers the position of both components
    relative to the BD, at $\sim$ $0\farcs08$ (in opposing directions). 
A position velocity diagram of the line region supports the
 spectro-astrometric results. The \Ha\ and HeI$\lambda$6678 lines were also analysed. 
 These line regions are not offset with respect to the
  continuum ruling out the presence of spectro-astrometric artifacts and underlining 
  the validity of the \Ox\ results. The low radial velocity 
  of the outflow, and relatively large offsets, are consistent with 
2MASS1207-3932 having a near edge-on disk, as proposed by Scholz05 et al.
 2MASS1207-3932 is now the smallest mass galactic object known to 
 drive an outflow. The age of the TW Hydrae Association 
 ($\sim$ 8 Myr) also makes this one of the oldest objects with
  a resolved jet. This discovery not only highlights the robustness 
  of the outflow mechanism over an enormous range of masses
   but also suggests that it may even be feasible for 
   young giant planets with accretion disks to drive outflows. 

\end{abstract}





\section{Introduction}
Recent studies have highlighted the numerous ways in which young brown dwarfs (BDs)
 bear a striking resemblance to classical T Tauri stars (CTTSs) \citep{Jay03a, Natta04, Whelan05, Scholz06}. 
This comparison not only provides information relevant to theories of BD formation, but also allows the validity of the accretion/ejection models for the formation of solar-mass stars, 
to be tested in the sub-stellar domain. Emphasis has been  
 on probing accretion \citep{Jay03b, Natta01}. While such studies point to T Tauri-like infall in BDs, 
relatively little is known about their outflow activity. Outflows in CTTSs are directly related to magnetospheric infall
 \citep{Hartigan94, Konigl00}, 
hence it is feasible that BDs demonstrating strong accretion will drive jets. 

The H$\alpha$\ 10$\%$ line width criterion (full width of the line at 10$\%$ of the peak emission) used to distinguish accreting T Tauri stars 
from those where the line emission is primarily due to chromospheric activity, has been successfully applied to BDs \citep{Jay03a}. A BD with a H$\alpha$\ 10$\%$ width $>$ 200 kms$^{-1}$ is accepted 
as a strong accretor. High resolution spectroscopic observations subsequently found that many of the accreting BDs also have  
forbidden emission lines (FELs) \citep{Fernandez01, Muzerolle03}. The FELs of CTTSs are important outflow tracers and their discovery 
provided the first hint that accreting BDs could also power outflows. Direct imaging has proven 
very difficult as expected on theoretical grounds \citep{Masciadri04}. 
An alternative approach adopted by \cite{Whelan05} is to search for sub-stellar outflows through the spectro-astrometric analysis of FELs. 
Using this method the first BD ouflow, driven by $\rho$-Oph 102, was discovered. Here the detection of a faint bipolar jet from a second strongly accreting BD, 2MASS1207-3932 
(hereafter 2MASS1207) using the ESO VLT, is reported. These observations are part of an important on-going study 
to search for outflows from BDs.

 That such outflows are present is physically reasonable since it seems unlikely that the outflow mechanism has anything to do with the eventual triggering of fusion in the core. 
 Moreover jets from the very low mass stars Par-Lup3-4 
\citep{Fernandez05} and ESO-H$\alpha$ 574 \citep{Comeron06} have been found recently. In addition a molecular outflow from the L1014 dense 
low luminosity core and possible BD precursor has been found by \cite{Bourke05}.

 \section{Target, Observations and Analysis}
 2MASS1207 was identified by \cite{Gizis02} and \cite{Mohanty03} as a sub-stellar member of the TW Hydrae Association.  
Given the age of the association ($\sim$ 8 Myr), the evolutionary tracks of \cite
{Chabrier00} were then used to estimate a mass of $\sim$ 35 M$_{jup}$ for this BD.  More recently the mass has been revised downwards to
 24M$_{jup}$ 
$\pm$ 6M$_{jup}$ \citep{Chauvin05, Mohanty06}. 
The nature of this BD as a strong accretor was clearly evident from early observations. \cite{Mohanty03} reported bright asymmetric \Ha\ emission, with a full width 
at 10$\%$ of the peak flux, 
of $>$ 200 \km . The new optical spectra presented here, lead to a value of $\sim$ 320 \km. \cite{Mohanty05} 
were the first to observe the \Ox\ line in the spectrum of 2MASS1207 suggesting it could be driving an outflow. 

2MASS1207 is known for its line variability. \cite{Scholz05} studied its \Ha\ emission over a period of 6 weeks and reported a change by a factor of 5-10 in the accretion 
rate during that time.  They also used the shape and variability of the \Ha\ line to infer a near edge-on orientation for the accretion disk. 
The prominent red-shifted absorption feature in the \Ha\ line (see Figure 1b) is strongly dependent on the inclination $\it {i}$, between the rotational axis and the line of sight \citep{Hartmann04}. The absorption feature should only be observed when the disk is seen close to edge-on i.e. 
 $\it {i}$ $\geq$ 60$^{\circ}$.  Refer to \cite{Scholz05} and \cite{Scholz06} for further details.  
 Such a geometry suggests that an outflow, will be in the plane of the sky. Lastly 2MASS1207 is a binary with a planetary mass 
 companion \citep{Chauvin04, Chauvin05}. To summarise 2MASS1207 is an ideal candidate to include in our 
 search for outflows driven by young BDs. It is a strong accretor with known \Ox\ emission. 
 
 High resolution (R=40,000) cross-dispersed spectra of 2MASS1207 (CD3 disperser, spectral range 4810-6740 \AA) were obtained with UV-Visual Echelle Spectrometer (UVES) 
  \citep{Dekker00} on the ESO VLT UT2, on May 16 2006. The results presented in this letter represent the sum of two 45 min exposures with a 1\arcsec\ slit. The seeing 
  was 0\farcs6 and the spatial sampling was 0\farcs182 per pixel. The data was reduced
using standard IRAF routines, and the 2D spectra obtained were smoothed using an elliptical Gaussian filter of FWHM 0.12 \AA $\times$ 0\farcs22. The Gaussian smoothing improved the signal to noise 
and effectively decreased the spectral and spatial resolution to R $\sim$ 8000 and 0\farcs7 respectively. The night sky lines were then removed. All velocities are quoted 
with respect to the systemic velocity of 2MASS1207 which is measured from the  LiI$\lambda$6708 line to be + 2\km, in the LSR frame. 
The final stage in the analysis was to apply the technique of spectro-astrometry to the smoothed spectra. 

For the majority of protostars with jets, the FEL regions of their spectra are very obviously spatially extended, particularly at high velocities, e.g. the FELs of the CTTS DG Tau
 trace a $\sim$ 12\arcsec\ knotty jet \citep{Whelan04}. When however the emission is very compact and close to the source, specialised methods have to be used to probe its nature. 
 In particular spectro-astrometry can investigate through Gaussian fitting, the 
 offset of emission line features smaller than the seeing disc of the observation.  
The result of this technique is an offset-velocity diagram with the displacement of a spectral feature, e.g. the \Ox, line shown as 
a function of velocity and relative to the continuum centroid of the object \citep{Bailey98, Takami01, Whelan04, Whelan05}. 
Figure 1 shows the results of the spectro-astrometry and see \cite{Whelan05} for a thorough discussion of the method.

\begin{figure}
\includegraphics[height=13cm,width=9cm]{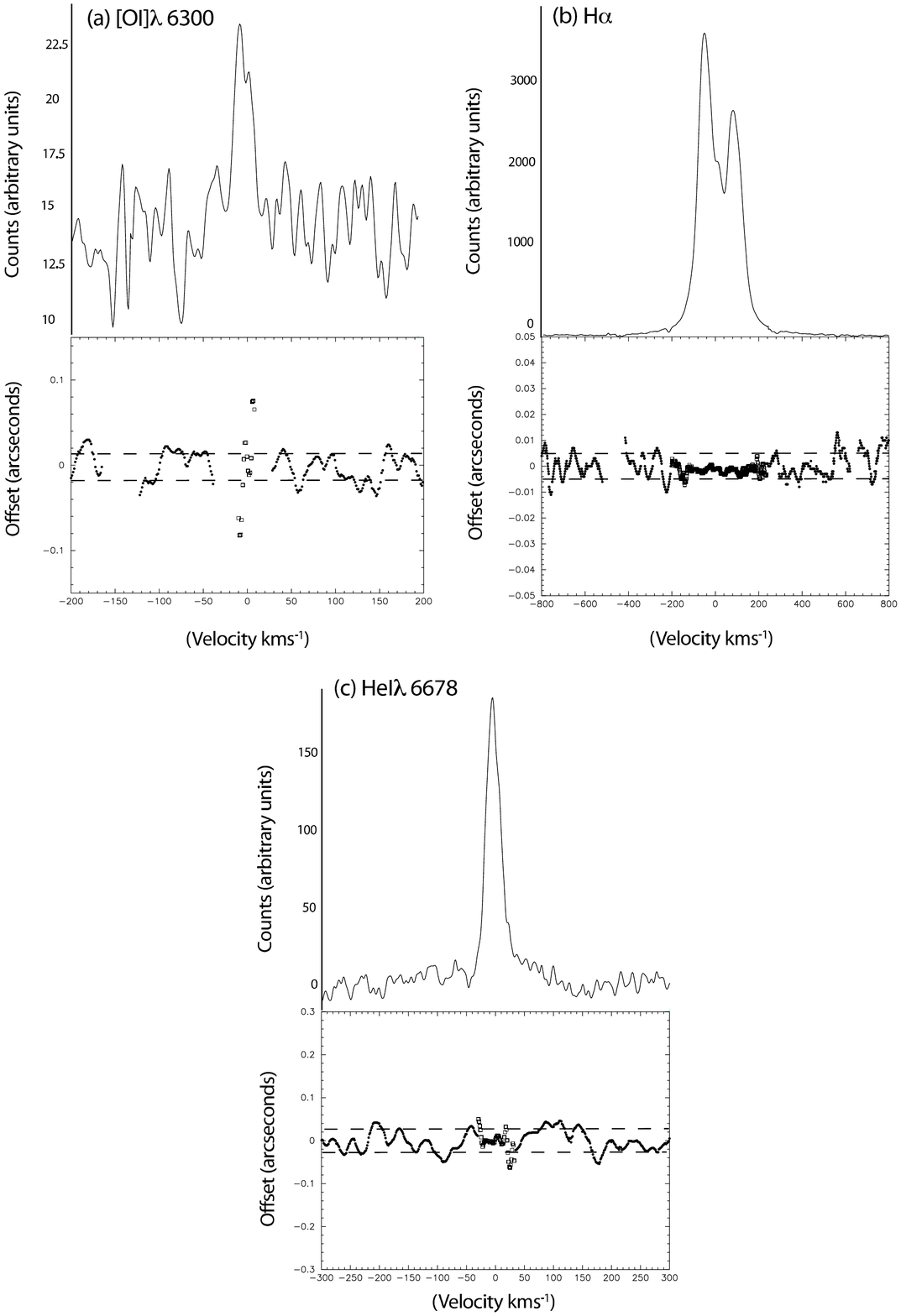}
\caption{Offset velocity diagrams in the vicinity of the \Ox, \Ha\ and He I lines. The line profiles are an average of three pixel rows in the spatial direction
  centred on the continuum. Note the continuum has not been 
 subtracted.  The green dashed lines delineate the $\pm$ 1 $\sigma$ 
 error envelope for the centroid position of the continuum. As explained in the text the line and continuum centroid errors are the same in all cases. 
 In particular for \Ha\ 
 the spectrum is smoothed so that error in the centroid position of the wing emission is comparable to that for the continuum. The error decreases for the 
 line peak region as is 
 clear in (b). The red-shifted absorption feature in the \Ha\ line is a signature of infall. The bipolar offset in the \Ox\ reveals the presence of an outflow. 
 No offsets 
 are measured in the \Ha\ and He I  lines as expected ruling out the possibility of spectro-astrometric artifacts. Lastly the gaps in the position of the continuum 
 blueward of the \Ox\ and \Ha\ lines, mark the position of night sky lines. The night sky lines have been reomved from the line profiles and as 
 found by \cite{Whelan05} the presence or not of these sky lines has no effect on the spectro-astrometric results.}

\end{figure}

\section{Results}
 
The \OI\ lines are the only  ``\,traditional\," FEL tracers seen in the UVES spectrum. A position velocity  (PV) diagram of the smoothed \Ox\ spectrum is presented in Figure 2. 
Blue and red-shifted emission is observed at velocities of $\sim$ -8 \km\ and $\sim$ + 4\km\ and a relative displacement between the 
two parts of the line is apparent. Also the relatively very small radial velocity 
of the outflow is consistent with the near edge-on disk hypothesis for 2MASS1207. We observe the [OI]$\lambda$6300 line at the same velocity as 
reported by \cite{Mohanty05}. However due to the use 
of a larger telescope (8m as opposed to 6.5m) and a longer exposure time we retrieved more detailed kinematic information.  
Note also that \cite{Scholz05} failed to detect forbidden emission in the spectrum of this BD. Again this is due to the use of a smaller telescope and 
the larger spectral resolution of their observations (R $\sim$ 60,000). The [OI] night sky line has been removed here and there is emission (at $\sim$ -35\km) on the farside of 
the night sky line position. This is probably also associated with the blue-shifted outflow. 

The results of a spectro-astrometric analysis of the smoothed spectrum is shown in Figure 1(a). The first step in this analysis is to map the position of the BD through 
Gaussian fitting of the spatial profile of the continuum. As the optical continuum emission from a BD, is by definition much weaker than 
for a CTTS say, it is necessary to bin or smooth the BD continuum in order to increase the spectro-astrometric accuracy. The error in the spectro-astrometric analysis depends strongly on the S/N and is given by 
$\sigma_{centroid}$ = $\frac{seeing(mas)}{2.3548\sqrt{\it N_p}}$ where ${\it N_{p}}$ is the number of detected photons. Once the continuum is mapped it is subtracted and the spatial profile of the 
pure line emission analysed. An approach first adopted by \cite{Whelan05} was to bin the continuum and line emission in such a way that as to ensure that a comparable number of photons are sampled at each point. In other words the spectro-astrometric accuracy achieved in 
the line and continuum is comparable. 

Here in the same way the continuum and \Ox\ line emission are smoothed so that the spectro-astrometric error in both regions are the same.
                        The blue and red-shifted 
components to the \Ox\ line are found to be 
offset in opposing directions to an absolute distance of $\sim$ 80 mas (see Figure 1a). Hence it is clear that the \Ox\ emission originates in a faint 
bipolar outflow driven by 2MASS1207. The 1-$\sigma$ error in the measurements, of 18 mas is marked, by the dashed line in Figure 1a. 
Also as in \cite{Whelan05} the presence or not of the \Ox\ 
night sky line was found to have no effect on the spectro-astrometric results. 

 \begin{figure}
 \includegraphics[height=10cm,width=7.5cm]{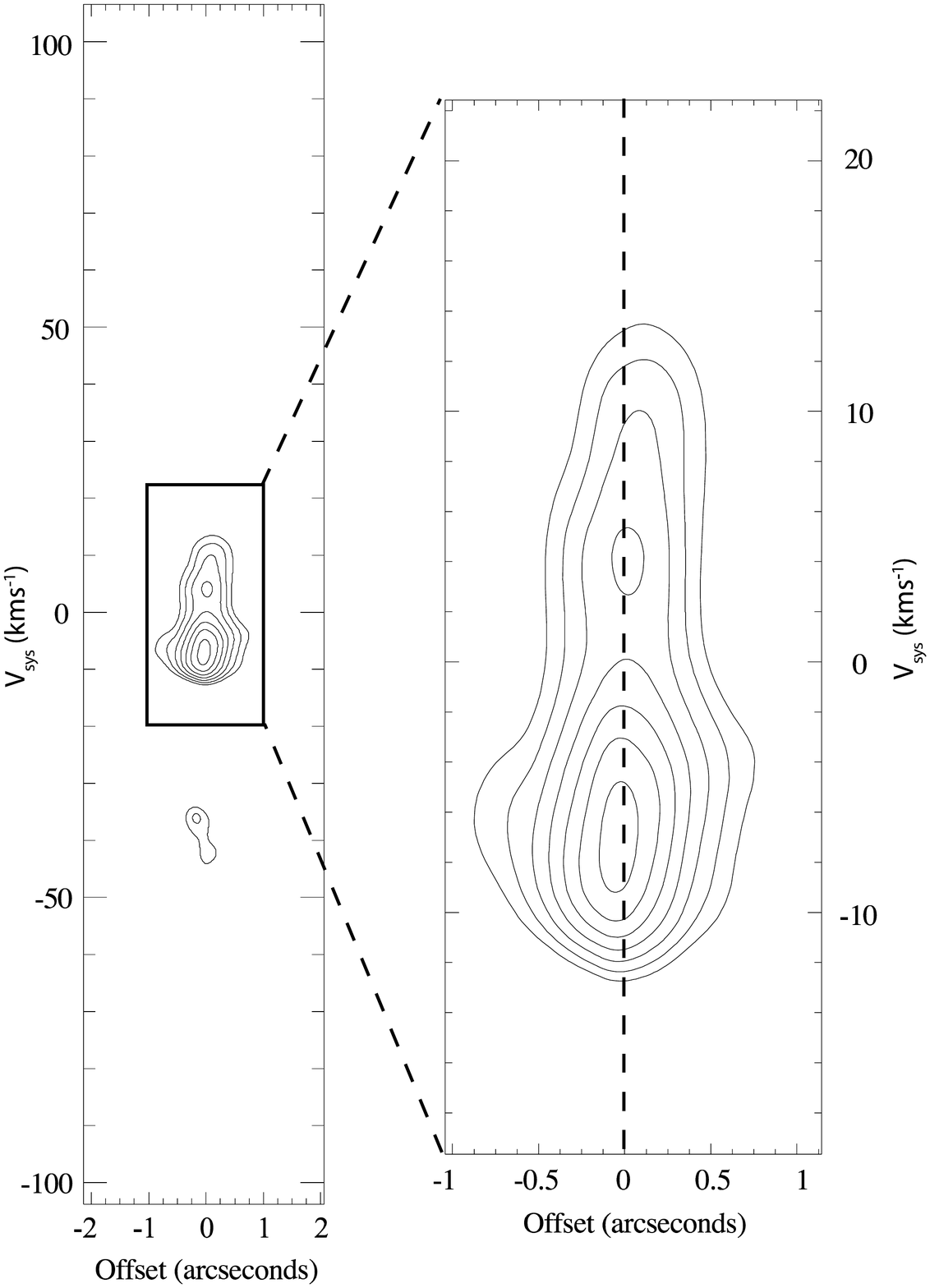}
  \caption{Continuum subtracted Position Velocity diagram of the \Ox\ line in 2MASS1207. Here the spectrum was smoothed in both the spectral and spatial directions using an 
 elliptical Gaussian function of FWHM 0.12 \AA $\times$ 0\farcs22. 
 The contours begin at 3 times the r.m.s noise and increase in intervals of the r.m.s noise. The zero spatial offset line is measured from an accurate mapping of the continuum position 
 using spectro-astrometry.
The red and blue-shifted components to the line are obvious and an opposing offset in these components is suggested. This offset is recovered using spectro-astrometry (see Figure 1). 
The \Ox\ night sky line has been removed from the spectrum and the feature blueward of this sky line, at $\sim$ -35 \km\ is probably also 
 part of the blue-shifted outflow.}
\end{figure}

Other lines of interest are [OI]$\lambda$6363, H$\alpha$ and HeI$\lambda$6678. The [OI]$\lambda$6363 is $\sim$ 3 times fainter than the \Ox\ 
line and hence is just below the detection 
limit in the raw spectrum. By smoothing the spectrum using a large gaussian filter (FWHM in dispersion direction = 0.35 \AA) the [OI]$\lambda$6363 is revealed at a similar velocity to the 
\Ox\ line, although detailed kinematic data is lost. The primary origin of the \Ha\ and HeI$\lambda$6678 lines is in the accretion flow. As this occurs on a very small scale we 
expect to measure no spectro-astrometric offset and
 as can be seen in Figure 1, none is found ruling out the presence of spectro-astrometric artifacts
\citep{Brannigan06}. 
The \Ha\ line is often seen in CTTS to have an outflow component. However at these scales any outflow component would certainly be extrememly weak relative 
to the bulk of the 
\Ha\ emission and therefore again very difficult to detect.    As 
the \Ha\ line is much brighter than any other part of the spectrum,  and as any outflow signature would be found in the wings, the line and the continuum 
are smoothed in such a way that the error in the continuum and the line-wings are comparable. All that is seen is a tightening of the \Ha\ 
position (due to the higher S/N) as one moves towards the peak of 
the line (Figure 1b). 

\begin{table}
\begin{tabular}{|llll|}       
 \hline\hline 
 Source                                   &2MASS1207     &$\rho$-Oph 102    &LS-RCrA 1          
 \\ 
\hline
Mass (M$_{jup}$)          &24                  &60                   &35              
\\
$\dot{M}_{acc}$ (\Msun$yr^{-1}$)  &10$^{-11}$   &1.25$\times10^{-9}$     &10$^{-9}$
\\
\Ox (\km, \arcsec)       &+4/-8,+0.08/-0.08  &-40, 0.085   &-9, --                           
 \\
\SI (\km, \arcsec)       &--,--  &-45, 0.085 &-15,--                 
\\ 
\NII     &--,--  &-45,-- &-19,--
\\ 
 \hline  
\end{tabular}
\caption{Table comparing the mass (in Jupiter masses), mass accretion rates and typical FEL velocities of 2MASS1207, $\rho$-Oph 102 and LS RCrA 1.
The accretion rate for $\rho$-Oph 102 was derived using the \Ha\ emission line and is taken from \cite{Natta04}. 
For 2MASS1207 and LS RCrA 1 \cite{Mohanty05} used the CaII$\lambda$8662 line fluxes to infer the mass 
accretion rates. The mass estimate for LS-RCr A1 is taken from \cite{Barrado04}.}

\end{table}

\section{Scaling Down from T Tauri Jets}
While it is accepted that accretion activity in BDs is simply a scaled down version of infall in CTTSs \citep{Jay03a}, the presumption 
that BDs also drive scaled-down versions of T Tauri jets and outflows while credible,
has yet to be generally confirmed. 
Here the evidence 
gathered so far supporting the continuation 
of the CTT paradigm for outflow activity, into the young BD mass regime, is summarised. 
Table 1 presents a comparison between the three BDs whose high resolution optical spectra 
have been analysed by us to date. LS-RCrA 1 probably has the strongest FELs of any BD  \citep{Fernandez01} and it is very likely that 
these FELs are excited in an jet. 
Yet direct imaging and a preliminary spectro-astrometric study failed to confirm such an origin for these features \citep{Fernandez05, Whelan06}. 
In particular spectro-astrometry is difficult due to the faintness of this objhect which in turn is almost certainly caused by an edge-on disk \citep{Barrado04}.
The relatively low radial velocities of the FELs of LS-RCrA 1 support this picture. 
From Table 1 $\rho$-Oph 102, 2MASS1207 and LS-RCrA 1 appear to launch outflows of comparable velocities. 
Also for 2MASS1207 and $\rho$-Oph 102 the \Ox\ line emitting region is offset by similar distances from the BD. Hence 
these three sub-stellar objects present a consistent picture of outflow activity in BDs,  
but how do their FEL regions (FELRs) compare to those of CTTSs? 

The most notable features of the FELRs of CTTSs are their blue-shifted and often multi-component nature \citep{Hirth97, Davis03}. The lack 
of red-shifted FE is accounted for by the obscuring effect of the 
accretion disk. Only blue-shifted emission is observed for $\rho$-Oph 102 and the scale of the blue-shifted offset (0$\farcs$085) 
reported by \cite{Whelan05} for $\rho$-Oph 102 would suggest a minimum (projected) disk radius of 0\farcs1 ($\geq$ 15 AU 
at the distance to the $\rho$-Ophiuchi cloud) in order to hide any red-shifted component. For 2MASS1207 the fact that both blue and 
red-shifted FE is detected is consistent with the near edge-on disk hypothesis. 

Lastly \cite{Whelan05} discuss the distance from a typical BD at which the critical density (n$_{cr}$) for forbidden emission is 
expected to be reached, if it is assumed that BD jets are similar in temperature, line excitation and opening angle to T Tauri jets. 
This distance is derived by considering how n$_{cr}$ (2 x 10$^{6}$ cm$^{-3}$ for \Ox) is 
expected to scale with mass flux through the jet. Using the predictions of \cite{Masciadri04}, the distance for forbidden emission (and thus measured offsets) is 
estimated to be 3-10 times closer to a BD than a CTTS. The \Ox\ offsets measured for 2MASS1207 lie within this mass range. These offsets are maximised due to 
the probable edge-on disk geometry of 2MASS1207. In conclusion there is much evidence in existence to support the continuation T Tauri like outflow activity into the mass 
range of young BDs.

\section{Conclusions}
Our analysis of the high spatial resolution spectrum of 2MASS1207 clearly uncovers the existence of a 
bipolar, T-Tauri like outflow. This result is in agreement with the classification 
of 2MASS1207 as a strong accretor with forbidden emission and 
its proposed near edge-on disk geometry. As 2MASS1207 is only a $\sim$ 24M$_{jup}$ object it now ranks as  
the least massive galactic object known to 
drive a jet. Given the age of the TW Hydrae Association it is also one of the oldest galactic objects with a resolved jet. The age of 2MASS1207 highlights that 
outflows can persist for a relativley long time, even when an object is sub-stellar, and the derived mass accretion rates are relatively low. 
Overall this discovery is not only intriguing but important as it emphasises the 
robustness of the outflow mechanism over an enormous range of masses (up to 10$^{8-9}$\Msun in Active Galactic Nuclei) and it adds weight to the idea 
that at least, massive planets 
may also be outflow sources 
when they form. See \cite{Quillen98} for an intriguing discussion of the possibility of outflows driven by proto-Jovian planets.

\acknowledgements{The present work was supported in part by the European Community's Marie Curie Actions - Human Resource and Mobility within the 
JETSET (Jet Simulations, Experiments and Theory) network under contract MRTN-CT-2004 005592 and by Science Foundation Ireland (contract no. 04/BRG/P02741). 
This work is based on observations 
that were taken at the European Southern Observatory, Chile (077.C-0771)}

\end{document}